\documentclass[conference,a4paper]{APSIPA2025}
\usepackage{amsmath,amssymb,amsfonts,amsthm}
\usepackage{graphicx}
\usepackage{booktabs}
\usepackage{multirow}
\usepackage{threeparttable}
\usepackage{url,times, tabularx,gensymb}
\usepackage[backend=biber,style=ieee,]{biblatex}
\usepackage{geometry}
\usepackage{fancyhdr}

\fancypagestyle{firststyle}{
  \fancyhf{}
  \fancyhead[C]{}
}

\begin{document}

\title{Enhancing Stereo Sound Event Detection with BiMamba and Pretrained PSELDnet}

\author{
\authorblockN{
Wenmiao Gao\authorrefmark{1},
Han Yin\authorrefmark{2}
}

\authorblockA{
\authorrefmark{1}
Department of Electrical and Photonics Engineering, Denmark Technical University, Lyngby, Denmark \\
\authorrefmark{2}
School of Electrical Engineering, KAIST, Daejeon, Republic of Korea\\
E-mail: wenmiaogao@163.com, hanyin@kaist.ac.kr}
}

\maketitle
\thispagestyle{firststyle}
\pagestyle{fancy}

\begin{abstract}
Pre-training methods have greatly improved the performance of sound event localization and detection (SELD). However, existing Transformer-based models still face high computational cost. To solve this problem, we present a stereo SELD system using a pre-trained PSELDnet and a bidirectional Mamba sequence model. Specifically, we replace the Conformer module with a BiMamba module. We also use asymmetric convolutions to better capture the time and frequency relationships in the audio signal. Test results on the DCASE2025 Task 3 development dataset show that our method performs better than both the baseline and the original PSELDnet with a Conformer decoder. In addition, the proposed model costs fewer computing resources than the baselines. These results show that the BiMamba architecture is effective for solving key challenges in SELD tasks. The source code is publicly accessible at \url{https://github.com/alexandergwm/DCASE2025_TASK3_Stereo_PSELD_Mamba}. 
\end{abstract}

\section{Introduction}
Sound event localization and detection (SELD) aims to identify sound events from a set of known classes, follow their changes over time, and estimate their positions in space when they occur~\cite{Adavanne2018Jun}. This task is important in many real-world applications. For example, it is used in robotic hearing systems, human-computer interaction, and immersive audio experiences. These technologies rely on accurate sound understanding to interact better with the environment.

The SELD task plays a key role in audio understanding by detecting sound events and locating their sources. To support this field, Detection and Classification of
Acoustic Scenes and Events (DCASE) has featured SELD as a regular task since 2019, increasing its difficulty each year to reflect real conditions. Early datasets used clean, simulated audio. In 2022, real recordings were added. In 2023, 360-degree video was introduced, creating an audio-visual track. In 2024, distance estimation was included. The 2025 version uses stereo audio and perspective video to better match media content. These changes highlight SELD’s rising importance and complexity.

Deep learning has brought major progress to sound event detection (SED)~\cite{Heittola2013Dec,7096611,fmsg_dcase2024,wilddesed,ucil,xiao2024mixstyle} and direction of arrival (DOA) estimation~\cite{Schmidt1986Mar,xiao2025wheresvoicecomingcontinual}, clearly outperforming earlier signal processing and machine learning techniques. A key milestone came with SELDnet by Adavanne et al.~\cite{Adavanne2018Jun}, which was the first end-to-end model to detect and localize sound events at the same time. However, this method could not handle cases where the same type of sound occurred at different locations. To solve this, the EINV2 model~\cite{Cao} introduced permutation-invariant training with a track-wise output format. It used two separate branches for SED and DOA, plus an extra shared branch. While effective, this design added training complexity and increased computation due to the need to balance different loss functions. To simplify the task, the class-wise ACCDOA format~\cite{Shimada2020Oct} was proposed. It combined detection and localization into a single vector for each sound class, making SELD a single-target problem. Later, the ADPIT method was developed to better handle overlapping sounds of the same class. The MultiACCDOA output format~\cite{Shimada2021Oct}, which extends this approach, is now widely used as the standard baseline.

Despite better output formats from MultiACCDOA, SELD still faces challenges, especially with training data. While pre-training and techniques like Audio Channel Swapping (ACS)~\cite{Wang2021Jan} help, it remains difficult to gather large, spatially rich datasets. This issue is worse with stereo data from FOA, as generating realistic audio-video pairs is resource-intensive. Limited data often leads to overfitting and weak generalization. Architecture-wise, models like PSELDnet~\cite{Hu2024Nov} rely on powerful pre-trained backbones~\cite{Kong2019Dec,Koutini2021Oct,Chen2022Feb} but still use heavy transformer-based architectures or large CNN models, These high-parameter models increase computational complexity significantly. Also, many systems treat spatial and temporal features separately, limiting full scene understanding.

To address this, we consider Mamba~\cite{Gu2023Dec,Dao2024May}, a new architecture combining CNNs’ local modeling with Transformers’ long-range ability using a state-space approach. Mamba scales linearly with input length \((O(n))\), making it efficient for long audio sequences. Its ability to capture both short and long-term patterns fits SELD’s needs well. Recent studies confirm Mamba’s value. In speech separation and enhancement~\cite{Li2024Apr,xiao2025tfmambatimefrequencynetworksound,xlsrmamba,Zhang2024May,Mu2024Aug}, BiMamba has replaced BiLSTM and Multi-Head Self-Attention, achieving strong results with less complexity. Mu et al.~\cite{Mu2024Aug} also replaced the Conformer module with BiMamba based on the EINV2 framework, surpassing EINV2's performance in multi-task output SELD systems. These outcomes show that BiMamba can improve performance while remaining computational efficient.

Based on these strengths, we propose a new approach that integrates BiMamba into the PSELDnet framework. Specifically, we replace the Conformer decoder with a BiMamba block to better model spatiotemporal patterns in stereo input and reduce computing cost. We also introduce asymmetric convolutions to decouple time and frequency features, further reducing computational load. This design supports the needs of the DCASE2025 task, where generating large-scale spatial audio-visual data is not feasible, and stereo input is required. By fine-tuning the pre-trained PSELDnet models with this improved structure, we aim to boost SELD performance while keeping the system efficient. 

\section{Proposed Method}


Training SELD systems on large datasets and fine-tuning pre-trained models on specific downstream datasets has shown excellent transfer and generalization~\cite{Hu2024Nov}. This approach often uses the FSD50K dataset~\cite{Fonseca2025} to collect clean, single-source sound clips, which are then convolved with room impulse responses generated by the mirror source method~\cite{Allen1979Apr}, before being converted to FOA format~\cite{Rafaely}.

However, for tasks involving conventional videos with stereo audio, creating audio-video datasets is technically challenging and requires significant resources. This makes it difficult to train models directly on large paired datasets. Therefore, effectively using pre-trained models is essential for achieving strong performance. To solve this problem and make stereo inputs compatible with pre-trained FOA-based models, we apply ACN/SN3D conversion techniques, which transforms stereo signals—simulating Mid/Side technology—into pseudo-FOA format, enabling smooth adaptation to existing pre-trained models. Specifically, the binaural signals $L(n)$ and $R(n)$ are transformed into FOA components according to Equation (\ref{equ:1}):

\begin{equation}
\begin{aligned}
    W(n) &= \frac{L(n) + R(n)}{2} \\
    Y(n) &= \frac{L(n) - R(n)}{2} \\
    X(n) &= 0\\
    Z(n) &= 0
    \label{equ:1}
\end{aligned}
\end{equation}

where $W(n)$ represents the omnidirectional component, $Y(n)$ represents the left-right directional component, $X(n)$ represents the front-back directional component, and $Z(n)$ represents the up-down directional component. Since stereo signals inherently lack precise spatial information, $X(n)$ and $Z(n)$ are set to zero. This four-channel pseudo-FOA representation is then processed to form the concatenation of log-mel spectrograms and intensity vectors as input.

\subsection{Theory of Mamba}

The Mamba architecture, developed from the Structured State Space Sequence (S4) model~\cite{Gu2021Oct,xiao2025tfmambatimefrequencynetworksound}, combines the strengths of CNNs and RNNs in a unified framework. It uses CNNs’ ability for parallel computation during training, while leveraging RNN-like temporal modeling during inference. This design helps the model efficiently capture both local and long-term dependencies in audio signals. Additionally, Mamba introduces a state-selection mechanism, which allows the network to focus on or ignore specific parts of the input sequence. This selective attention is essential for accurately identifying overlapping sound events, a common challenge in real-world SELD tasks.

  Specifically, inspired by continuous linear time-invariant systems in signal processing and control systems, it transforms the input sequence $x(t) \in \mathbb{R}$ to the output sequence $y(t) \in \mathbb{R}$ using higher dimensional hidden states $h(t) \in \mathbb{R}^{N \times 1}$, which can be written as follows:
\begin{equation}
\begin{aligned}
&    \mathbf{h}'(t) = \mathbf{A}\mathbf{h}(t) + \mathbf{B} x(t) \\
 &   y(t) = \mathbf{C}^T \mathbf{h}'(t) + \mathbf{D}x(t)
    \label{equ: 1}
\end{aligned}
\end{equation}
where $\mathbf{A} \in R^{N\times N}$, $\mathbf{B} \in R^{N \times 1}$,$\mathbf{C} \in R^{N \times 1}$ , and $\mathbf{D}$ represent the state transition matrix, the input projection matrix, the output projection matrix, and the skip connection matrix respectively. In practice, to handle discrete sequences, discretization of the SSM is necessary. Using the Zero Order Holding method, we introduce a time step $\Delta$ to sample the continuous matrices $\mathbf{A}$ and $\mathbf{B}$, obtaining discrete representations $\bar{\mathbf{A}}$ and $\bar{\mathbf{B}}$, as follows:

\begin{equation}
\begin{aligned}
    & \bar{\mathbf{A}} = \exp(\Delta \mathbf{A})\\
    &\bar{\mathbf{B}} = (\Delta \mathbf{A})^{-1}(\exp \Delta \mathbf{A} - \mathbf{I}) \cdot \Delta \mathbf{B}
    \label{equ:2}
\end{aligned}
\end{equation}
and the discretized structured SSM are as follows:
\begin{equation}
\begin{aligned}
    & \mathbf{h}_k = \bar{\mathbf{A}}\mathbf{h}_{k-1} + \bar{\mathbf{B}}x_k
    \\
    & y_k = \mathbf{C}^T \mathbf{h}_k
    \label{equ: 4}
 \end{aligned}
\end{equation}


The discretized parameters in Mamba vary over time through a selective state space modeling (SSM) approach, which works similarly to gating mechanisms in RNNs. This allows the model to decide at each time step whether to focus on or ignore certain input features, improving its ability to process information effectively. Building on this idea, Mamba2 introduces the SSD module~\cite{Dao2024May}, which further decomposes state evolution into two processes: intra-chunk parallel computation and inter-chunk recurrence. This design enables linear computational complexity while maintaining high accuracy when modeling long audio sequences, making it well-suited for demanding SELD tasks.

\subsection{Overall Pipeline}




Our proposed model builds on the CNN14-Conformer backbone from PSELDnet, using the pre-trained CNN14 encoder to extract strong feature representations. The model produces predictions in the Multi-ACCDOA (MACCDOA) format. To enhance temporal and spatial modeling, we replace the original Conformer blocks with BiMamba blocks, which are further improved by asymmetric convolutional layers.

\begin{figure}[]
    \centering 
    \includegraphics[width=1.0\columnwidth]{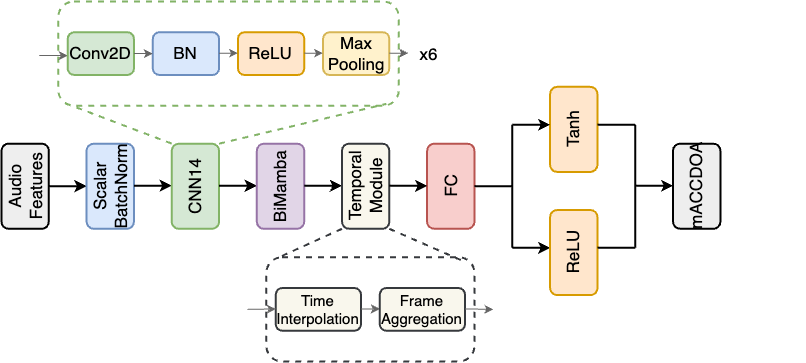}
    \caption{The overall architecture of the proposed system.}
    \label{fig:pipeline}
    \vspace*{-10pt} 
\end{figure}

As shown in Figure~\ref{fig:pipeline}, the overall architecture includes the pre-trained CNN14 block as the encoder and BiMamba blocks as the decoder. To resolve the temporal resolution mismatch caused by the multiple pooling layers in CNN14, we add a Temporal Module before the fully connected (FC) layers. This module ensures the model’s output aligns with the temporal resolution required by the labels.

The Temporal Module has two main components: Time Interpolation and Frame Aggregation. Time Interpolation restores temporal resolution lost through CNN14’s pooling layers, while Frame Aggregation adjusts the frame alignment to match the target label resolution. Finally, the model applies activation functions suited to each output type: Tanh is used for Cartesian coordinate outputs, and ReLU is applied to distance predictions. The final output, in Multi-ACCDOA format, provides comprehensive spatial and temporal localization of sound events. Specifically, the CNN14 encoder backbone consists of six VGG-style convolutional blocks. Each block includes \(3\times3\) convolution layers, followed by batch normalization, max-pooling, and ReLU activation functions. This hierarchical structure allows the model to progressively extract detailed spectral-temporal features from the input audio. As the network goes deeper, each block captures more abstract and complex acoustic patterns, enabling the model to build strong feature representations for sound event localization and detection.


For the decoder, our BiMamba module extends the original Mamba architecture by introducing a bidirectional processing mechanism. This design allows the model to capture both forward and backward temporal dependencies, overcoming the causal limitations found in traditional Mamba models. By processing information in both directions, the decoder gains a more complete understanding of the temporal context within audio sequences.

\begin{figure}[t]
    \centering
    \includegraphics[width=0.8\columnwidth]{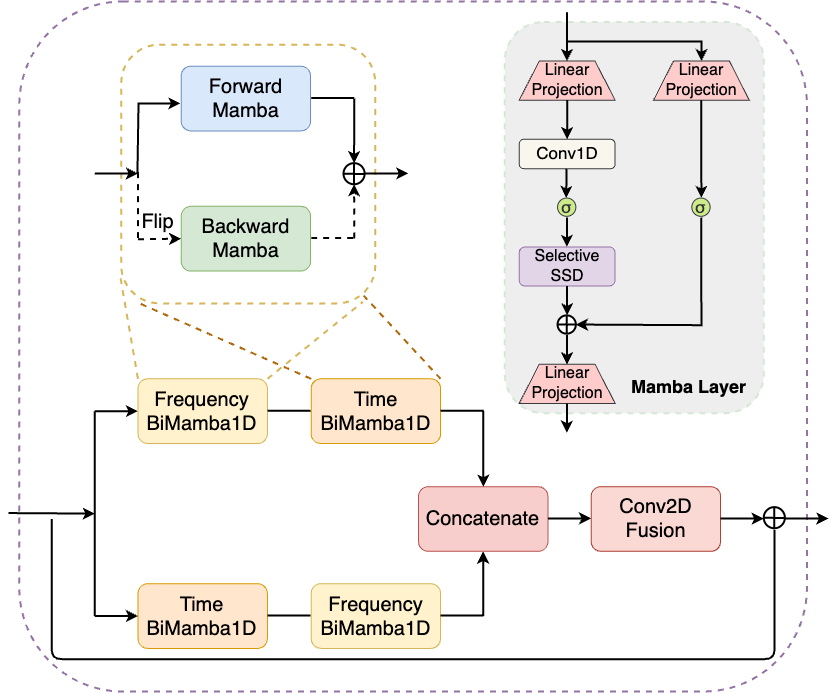}
    \caption{The detailed structure of the BiMamba2DAC Block.}
    \label{fig:bimamba_block}
\end{figure}

A key innovation in our approach is the use of asymmetric convolution, which separately processes the temporal and frequency dimensions. This design decouples feature extraction into two specialized pathways: one focused on dynamic patterns over time, and the other on acoustic features across frequencies. Figure~\ref{fig:bimamba_block} shows the detailed structure of the BiMamba block with asymmetric convolution, where gray blocks indicate Mamba layer components. This decoupled approach offers several advantages. First, replacing traditional 2D convolutions with 1D convolutions reduces computational complexity while maintaining strong feature extraction capabilities. Second, the dual-pathway design captures complementary aspects of the input, enabling richer and more robust feature representations. Finally, independent processing of temporal and spectral dimensions allows for targeted optimization in each domain, improving the model’s ability to detect dynamic events and identify their acoustic characteristics accurately.



\section{Experiments}
\subsection{Benchmark Dataset}

The proposed method is evaluated on the public benchmark dataset \textbf{DCASE2025 Task3 Stereo SELD Dataset} \cite{Shimada2025Jun}, which is derived from the original \textbf{STARSS23} \cite{Politis2023Mar}. This dataset consists of 5-second audio segments sampled from the source data, where the original FOA audio recordings are converted to stereo audio based on fixed viewpoints. The dataset encompasses 13 distinct sound event categories for comprehensive evaluation of sound event localization and detection performance.

\subsection{Comparable Models}
For comparative evaluation, we benchmark our proposed method against several models. The comparison includes the original \textbf{CNN14-Conformer} model \cite{Kong2019Dec} as the foundational baseline, the \textbf{HTS-AT} model \cite{Chen2022Feb} based on pure Transformer architecture, the \textbf{CNN14-ConBiMamba} \cite{Zhang2024May} and \textbf{CNN14-ConBiMambaAC} models, which replace the Multi-Head Self-Attention (MHSA) components in Conformer blocks with BiMamba and BiMambaAC mechanisms, respectively. The \textbf{CRNN} model \cite{Adavanne2018Jun} is also regarded as an additional reference.

\subsection{Model Configurations}
Audio signals are resampled to 24 kHz and processed using 64 mel-scale filters. STFT computation uses a 20ms hop length (480 samples) and 40ms window length (960 samples). The model employs 2 BiMamba2DAC blocks with recommended parameters: $d_{state} = 64$, $d_{conv} = 4$, and $expand = 2$.

For training, all models use the Adam optimizer with ReduceLROnPlateau scheduling, monitoring validation F1-score and reducing the learning rate by half after 5 stagnant epochs. Models are trained for 120 epochs with full parameter fine-tuning. Training hyperparameters include a learning rate of 1e-4 and weight decay between 1e-4 and 5e-6, with batch sizes ranging from 32 to 256 depending on model size.

Evaluation follows the DCASE2025 Task 3 protocol, selecting the checkpoint with the highest validation location-dependent F1-score. Performance metrics include location-dependent F1-score at a 20° threshold ($F_{20°}$), DOA Error (DOAE), and Relative Distance Error (RDE). Model complexity is assessed by parameter count and MACs computed on 5-second audio segments sampled at 24 kHz.

\section{Results}

We evaluate our proposed BiMambaAC against several baseline models under two training scenarios: training from scratch and fine-tuning from pretrained PSELDnet weights.

\subsection{Training from Scratch}

Table \ref{tab:scratch_performance} compares different model architectures trained from scratch on the development set. Conformer achieves the highest $F_{20^\circ}$ score of 32.7\% with excellent directional accuracy (14.5° DOAE) but requires 210M parameters and suffers from poor distance estimation (47\% RDE). Our BiMambaAC delivers competitive performance with 32.1\% $F_{20^\circ}$ and comparable directional accuracy (14.8° DOAE) using only 76M parameters while maintaining significantly better distance localization (36\% RDE). HTS-AT shows 25.2\% $F_{20^\circ}$ with the most efficient parameter usage (28M), demonstrating a favorable accuracy-efficiency trade-off. The lightweight CRNN baseline establishes a lower bound at 22.8\% $F_{20^\circ}$.

\begin{table}[htbp]
\centering
\caption{Comparison of different model architectures on the development set trained from scratch}
\label{tab:scratch_performance}
\resizebox{\linewidth}{!}{
\begin{tabular}{lccccc}
\toprule
Model & Params & MACs & $F_{20^\circ} \uparrow$ & $DOAE \downarrow$ & $RDE \downarrow$ \\
\midrule
Conformer  & $210M$ & $4.69G$ & $\textbf{32.7\%}$ & $\textbf{14.5\degree}$ & $47\%$ \\
HTS-AT & $28M$ & $2.88G$ & $25.2\%$ & $18.0\degree$ & $40\%$ \\
CRNN & $0.7M$ & $57M$ & $22.8\%$ & $24.5\degree$ & $41\%$ \\
BiMambaAC  & $76M$ & $4.63G$ & 32.1\%& 14.8\degree & $\textbf{36\%}$ \\
\bottomrule
\end{tabular}}
\end{table}

\subsection{Fine-tuning from Pretrained PSELDnet}

Table \ref{tab:pretrained_performance} shows that pretrained PSELDnet initialization significantly improves performance across all models. Our BiMambaAC achieves the best overall performance with 39.6\% $F_{20^\circ}$ and 15.8° DOAE using only 76M parameters, outperforming BiMamba (+3.4\% $F_{20^\circ}$) while reducing parameters by 57\% and computation by 39\%.
The ablation study in Table \ref{tab:ablation_study} validates the importance of each architectural component. Removing bidirectional processing causes the largest performance drop (-8.6\% $F_{20^\circ}$), while eliminating asymmetric convolution reduces accuracy by 6.4\%. Although increasing $d_{state}$ from 64 to 128 slightly improves performance (+0.3\% $F_{20^\circ}$), it requires 6GB additional GPU memory during training, making the current configuration more practical for computational efficiency.

The ConBiMamba variants show mixed results. ConBiMamba exhibits poor distance localization (53\% RDE) despite substantial computational overhead (338M parameters), while ConBiMambaAC improves efficiency but underperforms BiMambaAC, suggesting potential incompatibility between additional convolutional layers and our asymmetric convolution design.

\begin{table}[htbp]
\centering
\caption{Comparison of different model architectures on the development set fine-tuned from pretrained PSELDnet}
\label{tab:pretrained_performance}
\resizebox{\linewidth}{!}{
\begin{tabular}{lccccc}
\toprule
Model & Params & MACs & $F_{20^\circ} \uparrow$ & $DOAE \downarrow$ & $RDE \downarrow$ \\
\midrule
Conformer  & $210M$ & $4.69G$ & $38.2\%$ & $15.9\degree$ & $33\%$ \\
HTS-AT & $28M$ & $2.88G$ & $35.1\%$ & $16.5\degree$ & $\textbf{30\%}$ \\
CRNN & $0.7M$ & $57M$ & $22.8\%$ & $24.5\degree$ & $41\%$ \\
BiMamba & $178M$ & $7.57G$ & $36.2\%$ & $16.6\degree$ & $33\%$ \\
ConBiMamba & $338M$ & $7.98G$ & $36.2\%$ & $16.9\degree$ & $53\%$ \\
ConBiMambaAC & $236M$ & $5.02G$ & $33.7\%$ & $16.3\degree$ & $39\%$ \\
BiMambaAC  & $76M$ & $4.63G$ & \textbf{39.6\%} & \textbf{15.8\degree} & 33\% \\
\bottomrule
\end{tabular}}
\end{table}

\begin{table}[htbp]
\centering
\caption{Ablation study of BiMambaAC components}
\label{tab:ablation_study}
\begin{tabular}{>{\raggedright}p{4cm}ccc}
\toprule
Architectural Variant & \textbf{$F_{20^\circ} \uparrow$} & DOAE $\downarrow$ & RDE $\downarrow$ \\ 
\midrule
\textbf{Proposed} & 39.6\%& 15.8$^\circ$ & 33\%\\
w/o bidirectional Mamba  & 31.0\% & 17.1$^\circ$ & 37\% \\
w/o Asymtric convolution & 33.2\% & 17.6$^\circ$ & 39\% \\
Increased States ($d_{state}=128$) & 39.9\% & 15.5$^\circ$ & 35\% \\
\bottomrule
\end{tabular}
\end{table}

Our BiMambaAC method achieved 9th place in DCASE 2025 Task 3 using only single-task training with official dataset and simple channel swapping augmentation. This demonstrates the effectiveness of pretrained model fine-tuning and asymmetric convolution design. As shown in Table \ref{tab:seld_comparison}, our approach outperforms ensemble-based methods while maintaining a simpler single-model architecture compared to higher-ranked methods that rely on external datasets and multiple augmentation strategies.
\begin{table}[htbp]
\centering
\caption{Comparison of SELD challenge submissions based on official evaluation dataset}
\label{tab:seld_comparison}
\resizebox{\linewidth}{!}{
\begin{tabular}{lcccccc}
\toprule
\textbf{Team} & \textbf{$F_{20^\circ} \uparrow$} & \textbf{$DOAE \downarrow$} & \textbf{$RDE \downarrow$} & \textbf{Aug kinds} & \textbf{Ext. Data} & \textbf{Mode} \\
\midrule
Wan\_XJU \cite{Wan_XJU_task3a_report} & 35.4\% & 18.6\degree & 34.9\% & 1 & 1 & Single \\
Zhao\_MITC-MG \cite{Zhao_MITC-MG_task3a_report} & 34.0\% & 16.8\degree & 36.6\% & 5 & 2 & Single \\
\textbf{Proposed} & 31.0\% & 17.4\degree & 40.1\% & 1 & --- & Single \\
Park\_KAIST \cite{Park_KAIST_task3a_report} & 30.3\% & 14.6\degree & 32.4\% & 2 & --- & Ensemble \\
Bahuguna\_UPF \cite{Bahuguna_UPF_task3a_report} & 28.8\% & 21.2\degree & 100\% & --- & 2 & Ensemble \\
\bottomrule
\end{tabular}}
\end{table}
\section{Conclusions}

This work presents a stereo SELD approach that combines PSELDnet pretraining with BiMamba sequence modeling. Results show that pre-trained models, especially HTS-AT, significantly improve F1-score and DOAE over the CRNN baseline. BiMamba-based models further enhance performance, confirming the strength of state space modeling for audio sequences. The BiMambaAC model, which integrates asymmetric convolution, achieves the best performance across all metrics with fewer parameters than larger models like ConBiMamba. These findings highlight the benefits of combining BiMamba and asymmetric convolution for accurate and efficient stereo SELD, offering a strong direction for future research.









\newpage
\footnotesize
\printbibliography

\end{document}